\documentclass[nofootinbib,twocolumn,showpacs,aps,epsf,amsfonts,amssymb,amsbsy]{revtex4}

\usepackage{graphics,bm}
\usepackage{epsfig}
\usepackage{graphicx}

\newcommand{\beq}{\begin{equation}}
\newcommand{\eeq}{\end{equation}}
\newcommand{\bqa}{\begin{eqnarray}}
\newcommand{\eqa}{\end{eqnarray}}

\def\sumint{\hbox{$\sum$}\!\!\!\!\!\!\int}
\def\square{\vcenter{\vbox{\hrule height.4pt
          \hbox{\vrule width.4pt height4pt
          \kern4pt\vrule width.3pt}\hrule height.4pt}}}

\begin{document}

\author{Jens O. Andersen} 
\email{andersen@tf.phys.no}
\affiliation{Department of Physics, 
Norwegian Institute of Science and Technology, N-7491 Trondheim, Norway}
\title{Pion and Kaon Condensation at Finite Temperature and Density}

\date{\today}

\begin{abstract}
In this paper, we study $O(2N)$-symmetric $\phi^4$-theory
at finite temperature and density using the 2PI-$1/N$ expansion.
As specific examples, we consider pion condensation at finite 
isospin chemical potential and kaon condensation 
at finite chemical potential for hyper charge and isospin charge.
We calculate the phase diagrams
and the quasiparticle masses for pions and kaons in the large-$N$ limit.
It is shown that the effective potential and the gap equation can be
renormalized by using local counterterms
for the coupling constant and mass parameter, which are independent of 
temperature and chemical potentials.

\end{abstract}
\pacs{11.15Bt, 04.25.Nx, 11.10Wx, 12.38Mh}

\maketitle


\section{Introduction}
$O(N)$-symmetric field theories are playing a significant role 
as effective quantum field theories in 
condensed matter and high-energy physics. For example, the 
nonrelativistic $O(2)$-symmetric model
is used to describe cold atomic Bose gases in the context of Bose-Einstein
condensation in harmonic traps~\cite{pit} and the
homogeneous dilute Bose gase~\cite{jens2}.
Similarly, the $O(4)$ (non)linear sigma model is used as a
low-energy effective theory for QCD. 
In massless two-flavor QCD, the global symmetry group is
$SU(2)_L\times SU(2)_R$ and one then takes advantage 
of the isomorphism between this group and $O(4)$.
This effective theory describes the three
pions and the sigma particle. Similarly, the 
$SU(2)_L\times SU(2)_R$-symmetric theory
is a toy model for the description of kaon condensates~\cite{kaon,kaon2}
in the color-flavor locked phase~\cite{frankie} of high-density QCD.

The thermodynamic functions for hot field theories can be 
calculated as a series in the the coupling constant by resumming 
certain classes of diagrams from all orders of perturbation theory.
However, it turns out that 
the expansion of these functions in powers of the coupling constant
breaks down at temperature
unless the coupling is tiny~\cite{birdie,krammer,review}.
In order to obtain more stable results, one has developed
and applied more advanced 
resummation techniques,
such as screened perturbation theory~\cite{karsch},
hard-thermal-loop resummation~\cite{abs1}
and resummations based on the two-particle irreducible (2PI)
effective action~\cite{bir,jm1,epet,jurgen}.
These approaches all improve the converge of the perturbative series of
the thermodynamic quantities. 

At large chemical potential and zero temperature, resummation of classes of
diagrams is also required~\cite{mcl,bal} to obtain 
the thermodynamic functions as a series expansion in the coupling.
However, the perturbative series may be better behaved at zero temperature
than at high temperature as the plasmon effects are different~\cite{robstar}.
In this case, and also at nonzero but low temperature ($T\leq \mu$), 
hard-dense-loop resummation has proven to be an effective way of reorganizing
the calculations~\cite{reb}. Very recently, Ipp {\it et al}~\cite{ipp} 
have presented
a new method for calculating the pressure in deconfined QCD at all values
of the temperature and the quark chemical potential, which reduces to
the known perturbative results in the appropriate limits.

The $1/N$ expansion in conjunction with e.g.
the 1PI or 2PI effective-action formalism is 
another nonperturbative resummation scheme which has been applied at finite
temperature~\cite{meyer,kirk,gertie,abw}. 
A leading-order calculation typically gives rise to a gap equation for a
mass parameter, which is straightforward to solve.
At next-to-leading order (NLO), the resulting equations
become harder to solve. 
Due to nontrivial renormalization issues, 
the NLO problem using the 1PI $1/N$-expansion, 
has been solved rigorously only very recently~\cite{abw}.
In particular, it was shown that the thermodynamic potential
can be renormalized in a temperature-independent way only at the
minimum.
In the case of the 2PI formalism there are additional complications
since the gap equation for the propagator
becomes nonlocal and nontrivial to solve.
However, the
2PI formalism is superior when it comes to the description of nonequilibruim
phenomena~\cite{gertie}.

Renormalization of expressions obtained in resummation schemes such as
the 2PI effective-action formalism is a nontrivial issue. 
In these approximations, one generally resums
only certain classes of diagrams from all orders of perturbation theory.
Since only a subset of diagrams from a given order in peturbation theory
is taken into account, standard proofs of perturbative renormalizability
do not apply and one is not guaranteed that ultraviolet divergences can be
systematically eliminated. Moreover, it is not obvious that these
divergences are independent of temperature.
Signficant progress on these issues has been made in recent years
starting with the papers by van Hees and Knoll~\cite{hees}.
They showed that the equation for the
two-point function in scalar $\phi^4$-theory
can be renormalized by introducing a finite number
of counterterms. A more systematic study of the renormalization issues
in scalar $\phi^4$-theory was carried out by Blaizot, Iancu and 
Reinosa~\cite{bir3}. In particular, they focused on the 
elimination of temperature-dependent divergences.
This is sufficient to calculate physical quantities such as the 
thermodynamic pressure, but only recently has it been shown 
by Berges {\it et al}~\cite{berges2} how to 
fix all the counterterms needed to calculate the proper vertices which
are encoded in the effective action. These counterterms are local and 
they are independent of temperature and chemical potential.

The idea of a Bose-Einstein condensate of pions or kaons in the core of
compact stars has a long history and the interest in such condensates is
due to their far-reaching consequences for these objects. For example,
the presence of a condensate  results in enhanced neutrino 
cooling~\cite{reddik}. 
Moreover, in contrast to hadronic matter in heavy-ion collisions, bulk matter
in compact stars must (on average) be electrically neutral and so a
neutrality constraint must be imposed.

In the present paper, we discuss pion and kaon condensation at finite
temperature and density using the 2PI-$1/N$ expansion.
Pion condensation and the phase diagram of two-flavor QCD have
been investigated 
using chiral perturbation theory~\cite{son1,split,kog1,loewe}, 
lattice QCD~\cite{kog2,gupta},
ladder QCD~\cite{ravag0}, 
chiral quark model~\cite{antal}, and 
Nambu Jona-Lasinio models in the mean-field 
approximation~\cite{ravag1,china,ebert1}.
In Ref.~\cite{ebert1}, the effects of imposing charge 
neutrality were studied at zero temperature.
There have also been some applications of the linear sigma model
at finite temperature and 
density in either the large-$N$ approximation~\cite{china} or
the Hartree approximation~\cite{petro}. In both cases, the 
quasiparticle masses were calculated, but 
renormalization issues were not discussed.
In the present paper, we generalize part of the finite-temperature calculations
of Lenaghan and Rischke~\cite{kirk} by including chemical potentials.
More specifically, we study pion and kaon condensation at finite temperature
and density using the 2PI $1/N$-expansion 
paying attention to renormalization issues.

The paper is organized as follows. In Sec.~II, we briefly discuss interacting
Bose gases at finite density. In Sec.~III, we consider pions at
finite temperature and isospin chemical potential. We calculate the 
phase diagram and the quasiparticle masses. In Sec.~IV, we investigate
kaon condensation in the presence of chemical potentials for
isospin and hypercharge. In Sec.~V, we summarize and conclude.

\section{Interacting Bose gas}
The Euclidean
Lagrangian for a Bose gas with $N$ species of massive charged scalars is
\bqa\nonumber
{\cal L}&=&
(\partial_{\mu}\Phi_i^{\dagger})(\partial_{\mu}\Phi_i)
-{H\over\sqrt{2}}\left(\Phi_1+\Phi_1^{\dagger}\right)
+m^2\Phi^{\dagger}_i\Phi_i
\\&& 
+{\lambda\over2N}\left(\Phi^{\dagger}_i\Phi_i\right)^2
\;,
\label{lag}
\eqa
where $i=1,2,...,N$ and 
$\Phi_i=(\phi_{2i-1}+i\phi_{2i})/\sqrt{2}$ is a complex field.
If $H=0$, the theory described by Eq.~(\ref{lag})
has $(2N-1)N$ conserved charges which equals the number
of generators of the group $O(2N)$. 
If $H$ is nonzero, the $O(2N)$ symmetry is explicitly broken down to 
$O(2N-1)$.
A gas of $N$ species of  
bosons can be characterized by the expectation values
of the different conserved charges in addition to the temperature.
For each conserved charge $Q_i$, one may introduce a nonzero
chemical potential $\mu_i$. However, it is possible to specify the 
expectation values of different charges only if they commute.
The maximum number of commuting charges is $N$~\cite{arthur} 
and these can 
be chosen as 
\bqa
Q_{i}&=&\int\;d^3x\,j_{i}^0\;,
\eqa
where the the current densities $j_{i}^{\mu}$ are
\bqa
j_i^{\mu}&=&\phi_{2i}\partial^{\mu}\phi_{2i-1}
-\phi_{2i-1}\partial^{\mu}\phi_{2i}
\eqa
The incorporation of a conserved charge $Q_i$ is done by making the substitution
\bqa
\partial_0\Phi_i\rightarrow\left(\partial_0-\mu_i\right)\Phi_i\;\\ 
\partial_0\Phi_i^{\dagger}\rightarrow\left(\partial_0+\mu_i\right)
\Phi_i^{\dagger}\;.
\eqa
in the Lagrangian~(\ref{lag}).
From the path-integral representation of the free energy density $\cal F$
\bqa
e^{-\beta V\cal F}&=&\int{\cal D}\Phi^*_i{\cal D}\Phi_i 
e^{-\int_0^{\beta}d\tau\int d^3x\cal L}\;,
\eqa
the expression for the charge density can be written as
\bqa
Q_i&=&-{\partial{\cal F}\over\partial\mu_i}\,.
\eqa
If we introduce $k$ chemical potentials, the full symmetry group is
broken down to $[O(2)]^k\times O(2N-2k)$.
If $m^2<0$, the $O(2N)$ symmetry is spontaneously broken down to
$O(2N-1)$. Even if $m^2>0$, the symmetry may be broken if 
one of the chemical potentials, $\mu_i$, 
is larger than a critical chemical potential $\mu_c=m$.
In that case, the $\langle0|\phi_{2i-1}|0\rangle\neq0$ and the corresponding
$O(2)$ symmetry is broken.

\section{Pions}
In this section, we discuss pions at finite chemical potential $\mu_I$
for (the third component of the) 
isospin charge, which corresponds to the current density 
$j^{\mu}\sim\phi_2\partial^{\mu}\phi_3-\phi_3\partial^{\mu}\phi_2$, and 
$N=2$ in Eq.~(\ref{lag}).
In the vacuum, the $O(4)$ symmetry is broken down
to $O(3)$ by a nonzero vacuum expectation value $\phi_0$ for $\phi_1$.
The symmetry is broken either spontanously in the chiral limit or explicitly if
$H\neq0$. 

We first introduce a nonzero expectation value $\rho_0$ 
for $\phi_2$ to allow for a charged pion condensate. The field
$\Phi_1$ is then written as
\bqa
\Phi_1&=&
{1\over\sqrt{2}}(\phi_0+i\rho_0+\phi_1+i\phi_2)\;,
\eqa
where $\phi_1$ and $\phi_2$ are quantum fluctuating fields.
The inverse tree-level propagator $D_0^{-1}(\omega_n,p)$ reads
\begin{widetext}
\bqa
D_0^{-1}(\omega_n,p)=
\left(\begin{array}{ccccc}
\omega_n^2+p^2+m_1^2&{\lambda\over N}\phi_0\rho_0&0&0&...\vspace{2mm}
\\
{\lambda\over N}\phi_0\rho_0&\omega_n^2+p^2+m_2^2&-2\mu_I\omega_n&0&...
\\
0&2\mu_I\omega_n&\omega_n^2+p^2+m_3^2&&
\\
0&0&0&\omega_n^2+p^2+m_4^2&... \\
0&0&0&0&\omega_n^2+p^2+m_4^2 \\
...&...&...&...&...

\end{array}\right)\;,
\label{d0}
\eqa
\end{widetext}
where $p=|{\bf p}|$, $\omega_n=2n\pi T$ is the nth Matsubara frequency, and 
the treel-level masses are
\bqa
\label{msig}
m_1^2&=&m^2+{3\lambda\over2N}\phi_0^2+{\lambda\over2N}\rho_0^2\;,\\
\label{mpi}
m_2^2&=&-\mu^2_I+m^2+{\lambda\over2N}\phi_0^2+{3\lambda\over2N}\rho_0^2\;,\\
m_3^2&=&-\mu^2_I+m^2+{\lambda\over2N}\phi_0^2+{\lambda\over2N}\rho_0^2\;,\\
m_4^2&=&m^2+{\lambda\over2N}\phi_0^2+{\lambda\over2N}\rho_0^2\;.
\label{2n}
\eqa
Note that the mass parameter $m^2$ is negative in the remainder of this
section, as we want to break chiral symmetry spontaneously 
in the chiral limit in the absence of a chemical potential.

The 2PI effective potential can be written as
\bqa\nonumber
\Gamma[\phi_0,\rho_0,D]&=&
{1\over2}m^2\left(\phi_0^2+\rho_0^2\right)
+{\lambda\over8N}\left(\phi_0^2+\rho_0^2\right)^2
\\ && \nonumber
-{1\over2}\mu_I^2\rho_0^2
-H\phi_0
+{1\over2}{\rm Tr}\ln D^{-1}
\\ &&
+{1\over2}{\rm Tr}D_0^{-1}D
+\Phi[D]\;,
\label{ea1}
\eqa
where $D$ is the exact propagator and
$\Phi[D]$ is the sum of all two-particle irreducible vacuum diagrams.
The trace is over space-time as well as field indices.
The expectation values $\phi_0$ and $\rho_0$ satisfy the
stationarity conditions
\bqa
\label{e1}
{\delta \Gamma[\phi_0,\rho_0,D]\over\delta\phi_0}&=&0\;,\\
{\delta \Gamma[\phi_0,\rho_0,D]\over\delta\rho_0}&=&0\;.
\label{e2}
\eqa
The exact propagator $D$ satisfies the variational equation
\bqa
{\delta\Gamma[\phi_0,\rho_0,D]\over\delta D}&=&0\;.
\label{var}
\eqa
Using Eq.~(\ref{var}) and $\Pi=D^{-1}-D_0^{-1}$, 
where $\Pi$ is the exact self-energy,
the variational equation can 
be written as
\bqa
\Pi(D)&=&2{\delta\Phi[D]\over\delta D}\,.
\eqa
The diagrams that contribute to the self-energy matrix $\Pi(\omega,p)$ 
are obtained
by cutting a line in the vacuum diagrams.

\begin{figure}[htb]
\includegraphics[width=2.2cm]{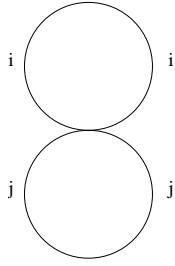}
\caption{
Feynman diagram contributing at leading order in $1/N$ to $\Phi[D]$.}
\label{lo}
\end{figure}

In Ref.~\cite{gertie}, the diagrams contributing to $\Phi[D]$ were classified
according to which order in the $1/N$ expansion they contribute.
The various terms can be written in terms of 
$O(N)$ invariants such as ${\rm Tr}(D^n)$  and ${\rm Tr}(\phi_0^2D^n)$, where $n$
is an integer. 
At leading order in $1/N$ the diagrams that contribute are shown
in Fig.~\ref{lo}. 
This diagram can be written as an $O(N)$-invariant term, 
$\Phi_{\rm LO}\sim\left({\rm Tr} D\right)^2$. 
The vertex gives a factor of $1/N$,
while each trace yields a factor  of $N$. It is therefore 
a leading-order contribution. 

\begin{figure}[htb]
\includegraphics[width=5.cm]{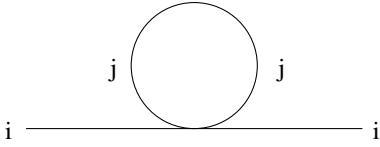}
\caption{
Feynman diagram contributing at leading order in $1/N$ to the self-energy 
$\Pi(\omega,p)$.}
\label{self}
\end{figure}

The diagram that contributes to the self-energy $\Pi$ to leading in $1/N$
is shown in Fig.~\ref{self}. 
The particle propagating in the loop is one of the $2N-3$ particles 
with tree-level masses given by Eq.~(\ref{2n}).

Their inverse propagator to leading order in $1/N$ can be written as
\bqa
D^{-1}(\omega_n,p)&=&P^2+m_4^2+\Pi_{\rm LO}(D)\;,
\label{fprop}
\eqa
where $P^2=\omega_n^2+p^2$
and $\Pi_{\rm LO}(D)$ is the tadpole in Fig.~(\ref{self}) calculated with
the full propagator $D$. Since $\Pi_{\rm LO}(D)$ is independent of the
external momentum, the propagator can be parametrized as
\bqa
D^{-1}(\omega_n,p)&=&P^2+M^2\;,
\label{para11}
\eqa
where $M$ is a constant. The leading order self energy can now be written
as
\bqa
\Pi_{\rm LO}&=&\lambda\sumint_Q{1\over Q^2+M^2}\;.
\eqa
Using the expression for $\Pi_{\rm LO}$ and the two expressions~(\ref{fprop}) 
and~(\ref{para11}) for the inverse propagator,
we obtain a gap equation for $M$:
\bqa
M^2&=&m_4^2+\lambda\sumint_Q{1\over Q^2+M^2}\;.
\label{gappi}
\eqa
The sum-integral is 
defined by
\bqa
\sumint_Q
&\equiv&\left({e^{\gamma}\Lambda^2\over4\pi}\right)^{\epsilon}
T\sum_{q_0=2\pi nT}\int_p{d^dq\over(2\pi)^d}
\;,
\eqa
where $d=3-2\epsilon$ and
$\Lambda$ is the renormalization scale associated with dimensional
regularization.
The integral over 3-momentum will be calculated
using dimensional regularization.
Inserting the parametrization~(\ref{para11}) of $D$ into the 2PI effective
potential~(\ref{ea1}) and differentiating with respect to $\phi_0$
and $\rho_0$, the equations~(\ref{e1}) and~(\ref{e2}) can be written as
\bqa\nonumber
0&=&m^2\phi_0-H+{\lambda\over2N}\phi_0(\phi_0^2+\rho_0^2)
\\&&
+{\lambda}\phi_0\sumint_Q{1\over Q^2+M^2}\;\;,
\label{g0}
\\
0&=&\nonumber
(m^2-\mu_I^2)\rho_0+{\lambda\over2N}\rho_0(\phi_0^2+\rho_0^2)
\\&&
+{\lambda}\rho_0\sumint_Q{1\over Q^2+M^2}\;.
\label{g22}
\eqa
Combining the gap equation~(\ref{gappi}) and Eq.~(\ref{g0}) gives
\bqa
H&=&M^2\phi_0\;.
\label{finull}
\eqa
Similarly, combining the gap equation~(\ref{gappi}) and Eq.~(\ref{g22}) gives
\bqa
\left(M^2-\mu^2\right)\rho_0&=&0\;.
\label{rhonull}
\eqa
We will study the various phases in the following subsections.

We next discuss the renormalization of the gap equation~(\ref{gappi}), 
which can be done by an iterative procedure given in Ref.~\cite{bir3}
(see also Ref.~\cite{patkos}).
By writing $M^2=m_4^2+\Pi_{\rm LO}$ and
expanding the right-hand side of Eq.~(\ref{gappi}) in powers of
$\Pi_{\rm LO}$, it is clear
that it resums the perturbative daisy and superdaisy diagrams:
\bqa\nonumber
M^2-m_4^2&=&\lambda\sumint_Q{1\over Q^2+m_4^2+\Pi_{\rm LO}}\\
&=&\nonumber
\lambda\sumint_Q{1\over Q^2+m_4^2}
-\Pi_{\rm LO}\lambda\sumint_Q{1\over(Q^2+m_4^2)^2}\\
&&+\Pi_{\rm LO}^2\lambda\sumint_Q{1\over(Q^2+m_4^2)^3}+...
\label{expanding}
\;.
\eqa

\begin{figure}[htb]
\includegraphics[width=7.cm]{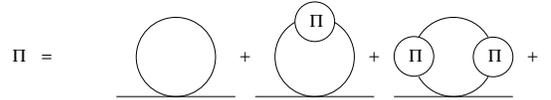}
\caption{
Diagrammatic interpretation of the gap equation~(\ref{gappi}) for
$\Pi_{\rm LO}$.}
\label{bub}
\end{figure}
This is illustrated diagrammatically in Fig~\ref{bub}.
Renormalization is carried by making the substitutions
\bqa
\label{c1}
m^2&\rightarrow&m^2+\sum_{n=1}^{\infty}\delta m^2_n\;, \\
\lambda&\rightarrow&\lambda+\sum_{n=1}^{\infty}\delta\lambda_n\;,
\label{c2}
\eqa
where $\delta m_n^2$ and $\delta\lambda_n$ are 
counterterms of order $\lambda^n$ and $\lambda^{n+1}$, respectively.
Similarly, we write the leading order self-energy
in a power series in $\lambda$:
\bqa
\Pi_{\rm LO}&=&\Pi_{\rm LO,1}+\Pi_{\rm LO,2}+...,
\label{c3}
\eqa
where the subscript indicates the power of $\lambda$. 
Inserting Eqs.~(\ref{c1})--(\ref{c3}) into equation~(\ref{expanding}),
we can determine $\Pi_{\rm LO,n}$,
$\delta m_n^2$ and $\delta\lambda_n$  
by iteration. 
The first iteration of the gap equation yields
\begin{widetext}
\bqa
\Pi_{\rm LO,1}&=&
\lambda\sumint_Q{1\over Q^2+m_4^2}
+\delta m^2_1+{\delta\lambda_1\over2N}\left(\phi_0^2+\rho^2_0\right) 
\;,
\label{gexp}
\eqa
\end{widetext}
where we have used that $\Pi_{\rm LO}=M^2-m_4^2$.
The next step is to calculate the sum-integral in Eq.~(\ref{gexp}).
The sum over Matsubara frequencies can be expressed as a contour integral
in the complex energy plane. By performing this integral, we obtain
\bqa\nonumber
\sumint_Q{1\over Q^2+M^2}&=&
\int_q{d^dq\over(2\pi)^d}{1\over\sqrt{Q^2+M^2}}
\\&&
\times\Bigg[{1\over2}+
{1\over e^{\sqrt{Q^2+M^2}/T}-1}
\Bigg]\;.
\eqa
The first term is divergent in the ultraviolet. Calculating it with 
dimensional regularization and expanding in $\epsilon$ through order
$\epsilon^0$, we obtain
\bqa\nonumber
\sumint_Q{1\over Q^2+M^2}&=&-
{M^2\over16\pi^2}\left({\Lambda^2\over M^2}\right)^{\epsilon}
\left[{1\over\epsilon}+1\right]
\\&&
\hspace{-1.7cm}
+{1\over2\pi^2}\int_0^{\infty}
{dq\,q^2\over\sqrt{q^2+M^2}}{1\over e^{\sqrt{q^2+M^2}/T}-1}
\label{sum1}
\;.
\eqa
The logarithimic divergence in the sum-integral in Eq.~(\ref{gappi})
shows up as a
pole in $\epsilon$, while the quadratic divergence is set to zero in
dimensional regularization.
Inserting the sum-integral~(\ref{sum1}) 
into Eq.~(\ref{gexp}), we obtain
\bqa\nonumber
\Pi_{\rm LO,1}&=&-{\lambda m_4^2\over16\pi^2\epsilon}
+\delta m^2_1+{\delta\lambda_1\over2N}\left(\phi_0^2+\rho^2_0\right) 
\\&&
+{\rm finite\;terms}\;.
\eqa
The divergences are removed by the counterterms
\bqa
\delta m_1^2={m^2\lambda\over16\pi^2\epsilon}\;,\hspace{1cm}
\delta\lambda_1={\lambda^2\over16\pi^2\epsilon}\;.
\eqa
This renormalization procedure can be carried out iteratively to all 
orders in $\lambda$ and this determines the counterterms
$\delta m_n^2$ and $\delta\lambda_n$ for $n>1$. The counterterms
$\delta m_n^2$ and $\delta\lambda_n$ are~\cite{bir3}
\bqa
\delta m^2_n={m^2\lambda^{n}\over\left(16\pi^2\epsilon\right)^n}\;,
\hspace{0.7cm}
\delta\lambda_n={\lambda^{n+1}\over\left(16\pi^2\epsilon\right)^n}\;.
\eqa
Summing the counterterms to all orders, Eqs.~(\ref{c1}) and~(\ref{c2})
can be written as
\bqa
m^2\rightarrow{m^2\over1-{\lambda\over16\pi^3\epsilon}}\;,
\hspace{1cm}
\lambda\rightarrow{\lambda\over1-{\lambda\over16\pi^3\epsilon}}\;.
\label{summing}
\eqa
Returning to the nonperturbative gap equation~(\ref{gappi}), it can be 
rewritten as
\begin{widetext}
\bqa\nonumber
{M^2\over\lambda}&=&
{m^2\over\lambda}
+{1\over2N}\left(\phi_0^2+\rho_0^2\right)
+\sumint_Q{1\over Q^2+M^2} \\ 
&=&{m^2\over\lambda}
+{1\over2N}\left(\phi_0^2+\rho_0^2\right)
-{M^2\over16\pi^2}\left({\Lambda^2\over M^2}\right)^{\epsilon}
\left[{1\over\epsilon}+1\right]
+{1\over2\pi^2}\int_0^{\infty}
{dq\,q^2\over\sqrt{q^2+M^2}}{1\over e^{\sqrt{q^2+M^2}/T}-1}
\label{sum3}
\;.
\eqa
\end{widetext}
Making the substitutions~(\ref{summing}), the renormalized gap
equation becomes
\bqa\nonumber
M^2&=&m^2
+{\lambda\over2N}\left(\phi_0^2+\rho_0^2\right)
-{\lambda M^2\over16\pi^2}\left(\ln{\Lambda^2\over M^2}+1\right)
\\ &&
+{\lambda\over2\pi^2}\int_0^{\infty}
{dq\,q^2\over\sqrt{q^2+M^2}}{1\over e^{\sqrt{q^2+M^2}/T}-1}\;.
\label{rengap}
\eqa


We finally discuss the renormalization of the effective potential
in the broken phase.
In the large-$N$ limit, $\Phi[D]$ is given by
\bqa
\Phi[D]&=&{1\over2}\lambda N\left(\sumint_Q{1\over P^2+M^2}\right)^2\;.
\label{expres}
\eqa
Using the expression~(\ref{expres}) and
inserting the gap equation~(\ref{gappi}) into the expression
for the effective potential~(\ref{ea1}), we obtain
\bqa\nonumber
\Gamma[\phi_0,\rho_0,D]&=&
{N\over2\lambda}\left(M^4-m^4\right)-{1\over2}\mu_I^2\rho_0^2
-H\phi_0\\&&
\hspace{-2.1cm}
-NM^2\sumint_Q{1\over P^2+M^2}
+N\sumint_Q\ln\left(P^2+M^2\right)
\label{effpotnew}
\;.
\eqa
The first sum-integral is given by Eq.~(\ref{sum1}), while the second
can be found by integrating the first with respect to $M^2$ and reads
\bqa\nonumber
\sumint_Q\ln\left(P^2+M^2\right)&=&
-{M^4\over32\pi^2}\left({\Lambda^2\over M^2}\right)^{\epsilon}
\left[{1\over\epsilon}+{3\over2}\right]
\\ &&
\hspace{-2.1cm}
+{T\over\pi^2}\int_0^{\infty}dq\;q^2\ln\left(1-e^{-\sqrt{q^2+M^2}/T}\right)\;.
\eqa
Inserting the expressions for the sum-integrals into~(\ref{effpotnew})
and making the substitutions~(\ref{summing}), 
we obtain the renormalized effective potential
\bqa\nonumber
\Gamma[\phi_0,\rho_0,D]&=&
{N\over2\lambda}\left(M^4-m^4\right)
-{1\over2}\mu_I^2\rho_0^2-H\phi_0
\\ &&\nonumber
+{NM^4\over32\pi^2}\left[\ln{\Lambda^2\over M^2}+{1\over2}
\right]
\\ &&\nonumber
-{NM^2\over2\pi^2}\int_0^{\infty}
{dq\;q^2\over\sqrt{q^2+M^2}}{1\over e^{\sqrt{q^2+M^2}/T}-1}
\\ &&
\hspace{-0.9cm}
+{NT\over\pi^2}\int_0^{\infty}dq\;q^2\ln\left(1-e^{-\sqrt{q^2+M^2}/T}\right)\;,
\eqa
where we have omitted a temperature-independent divergence proportional
to $m^4$. This corresponds to adding a vacuum counterterm $\Delta {\cal E}$
to the effective potential. We also note that the effective potential
can be renormalized in a similar manner in the symmetric phase.

\subsection{Chiral limit}
\label{climit}
In this subsection, we discuss the solutions to the gap equations and the
corresponding phases in the chiral limit, i.e. when $H=0$. 

First assume that the charged pion condensate  $\rho_0$ is nonzero.
Equation~(\ref{rhonull}) then immediately gives
\bqa
M^2&=&\mu_I^2\;.
\label{mmu}
\eqa
Equation~(\ref{finull}) then implies that the chiral condensate vanishes:
\bqa
\phi_0&=&0\;.
\label{null}
\eqa
Using Eqs.~(\ref{g22}),~(\ref{mmu}) and~(\ref{null}), the pion condensate 
reads
\bqa
\rho_0^2&=&
{2N\over\lambda}(\mu_I^2-m^2)
-2N\sumint_Q{1\over Q^2+\mu_I^2}\;.
\label{gaprho}
\eqa
The pole in $\epsilon$ on the right-hand side of this equation is 
removed by making the substitutions~(\ref{summing}) in analogy with the
renormalization of the gap equation~(\ref{gappi}).
After renormalization, the gap equation~(\ref{gaprho}) can be written as
\bqa\nonumber
\rho_0^2&=&{2N\over\lambda}\left(\mu_I^2-m^2\right)
-{N\mu_I^2\over8\pi^2}\left(\ln{\Lambda^2\over\mu_I^2}+1\right)
\\ &&
-{N\over\pi^2}\int_0^{\infty}{dq\;q^2\over\sqrt{q^2+\mu_I^2}}{1\over e^{\sqrt{q^2+\mu_I^2}/T}-1}
\;.
\label{rhot}
\eqa
As the temperature $T$ increases, 
Eq.~(\ref{rhot}) shows that $\rho_0$ goes to zero continuously and so
the phase transition is second order.
The critical temperature $T_c$ is given by the temperature for
which the right-hand side of Eq.~(\ref{rhot}) vanishes.
In the weak-coupling limit, where $\lambda\ll1$, 
we can obtain an analytic expression for 
$T_c$. In that case, $\mu_I$ can be neglected in the sum-integral 
in Eq.~(\ref{gaprho}) and its value is $NT^2/6$. This yields
the critical temperature
\bqa
T_c^2&=&{12\over\lambda}\left(\mu_I^2-m^2\right)\;,
\label{tcw}
\eqa
which is in agreement with the calculations of Kapusta~\cite{kapusta}.

The dispersion relations are obtained by analytic continuation
to Minkowski space, 
$\omega_n\rightarrow i\omega$, and then 
solving the equation ${\rm det}D(\omega,p)=0$.
For $T<T_c$, the dispersion relations are
\begin{widetext}
\bqa
\omega_1(p)&=&\sqrt{p^2+\mu_I^2}\;,\\
\omega_{2,3}(p)&=&
\sqrt{p^2+2\mu_I^2+{1\over2}\left(M_2^2+M^2_3\right)
\pm\sqrt{(M_2^2-M_3^2)^2+8\mu_I^2\left(2\mu^2_I+M_2^2+M_3^2\right)
+4\mu_I^2p^2}}\;,
\label{disp0} 
\\
\omega_{4}(p)&=&\sqrt{p^2+\mu_I^2}
\label{disp0p}
\;,
\eqa
\end{widetext}
where $M_i^2=m_i^2+\Pi_{\rm LO}$.
Note that the $\sigma$ and $\pi^0$ are identified with the original
fields $\phi_1$ and $\phi_4$, and therefore with
$\omega_1(p)$ and $\omega_4(p)$. On the other hand, the charged pions
$\pi^-$ and $\pi^+$ are given by linear combinations of $\phi_2$ and $\phi_3$
and are identified by $\omega_2(p)$ and $\omega_3(p)$.
In the minimum of the effective potential~(\ref{ea1}), we have
$M_2^2=2(\mu_I^2-\tilde{m}^2)$ 
and $M_3^2=0$, where $\tilde{m}^2=m^2+\Pi_{\rm LO}$.
The dispersion relations in Eq.~(\ref{disp0}) then reduce to
\bqa\nonumber
\omega_{2,3}(p)&=&\sqrt{p^2+3\mu_I^2
-\tilde{m}^2\pm\sqrt{(3\mu_I^2-\tilde{m}^2)^2+4\mu_I^2p^2}}\;.
\\&&
\label{disp1}
\eqa
From this equation, we see that the $\pi^+$ is a massless mode,
which in the long-wavelength limit behaves as
\bqa
\label{l1}
\omega_{3}(p)&=&\sqrt{{\mu^2_I-\tilde{m}^2\over3\mu^2_I-\tilde{m}^2}}\;p\;.
\eqa
This is consistent with the fact that the original $O(2)\times O(2)$
symmetry is broken down to $O(2)$ by the pion condensate.
The fact that the Goldstone mode is behaving linearly with $p$ makes
the system superfluid as the Landau criterion is satisfied~\cite{landau}.

In the normal phase, where $\phi_0=\rho_0=0$, the dispersion relations 
are
\bqa
\omega_1(p)&=&\sqrt{p^2+M^2}
\;,\\
\omega_{2,3}(p)&=&\sqrt{p^2+M^2}\pm\mu_I\;, \\
\omega_4(p)&=&\sqrt{p^2+M^2}\;,
\eqa
where $M$ is determined selfconsistently by solving the gap 
equation~(\ref{rengap}) evaluated at $\phi_0=\rho_0=0$:
\bqa\nonumber
M^2&=&m^2
-{\lambda M^2\over16\pi^2}\left(\ln{\Lambda^2\over M^2}+1\right)
\\ &&\!\!\!\!
+{\lambda\over2\pi^2}\int_0^{\infty}
{dq\,q^2\over\sqrt{q^2+M^2}}{1\over e^{\sqrt{q^2+M^2}/T}-1}\;.
\label{selfcon}
\eqa
\subsection{Physical point}
We next discuss the case where $H\neq0$. 
The solution to Eq.~(\ref{finull}) yields
\bqa
\phi_0&=&{H\over\mu_I^2}\;.
\label{finalt}
\eqa
Note that $\phi_0$ is nonzero and independent of the temperature.
Assuming a nonzero pion condensate,
the mass $M$ again is given by Eq.~(\ref{mmu}). 
Using Eqs.~(\ref{mmu}) and~(\ref{finalt}), the gap equation~(\ref{g22}) 
can be written as
\bqa
\rho_0^2&=&{2N\over\lambda}\left(\mu_I^2-m^2\right)-{H^2\over\mu_I^4}
-2N\sumint_Q{1\over Q^2+\mu_I^2}
\label{gapsup}
\;.
\eqa
The gap equation is renormalized in the same way as 
the gap equation in the chiral limit.

The expressions for the dispersion relations are very involved
and they are
not particularly illuminating. We therefore restrict ourselves to 
list the quasiparticle masses. These are found by solving the equation
$\det D(\omega,p=0)=0$ and read
\begin{widetext}
\bqa
\label{w1br}
\omega_1(p=0)=M_1=0\;, \\ 
\omega_{2,3}(p=0)=
M_{2,3}=\sqrt{
{7\over2}\mu_I^2-\tilde{m}^2
\pm{1\over2}\sqrt{(5\mu_I^2-2\tilde{m}^2)^2-12{\lambda\over N}
{H^2\over\mu_I^2}}}
\;, \\
\omega_4(p=0)=M_4=\mu_I\;,
\label{w4br}
\eqa
\end{widetext}
where $\tilde{m}^2=m^2+\Pi_{\rm LO}$.
The $\pi^+$ is again a massless Goldstone mode.
By expanding the equation for the dispersion relations around $p=0$, 
it can be shown that this mode
is linear in $p$ and hence the system is superfluid.
This mode reflects the breaking of the residual $O(2)$ symmetry and
is a conventional Goldstone mode.

We next discuss the system in the normal phase, where $\rho_0=0$.
Eliminating $\phi_0$ from the gap equation~(\ref{g0}),
one finds
\bqa
{2N\over\lambda}\left(M^2-m^2\right)&=&{H^2\over M^4}
+2N\sumint_Q{1\over Q^2+M^2}\;.
\label{gapnor}
\eqa
This equation is renormalized in the usual manner. The renormalized gap
equation determines the mass $M$ as a function of temperature.
We see that the equations~(\ref{gapsup}) and~(\ref{gapnor}) match onto each
other continously at the phase boundary, where $M=\mu_I$ as they should.
The dispersion relations in the symmetric phase
\bqa
\label{w1sym}
\omega_1(p)&=&\sqrt{p^2+M^2+{\lambda\over N}{H^2\over M^4}}\;, \\
\omega_{2,3}(p)&=&
\sqrt{p^2+M^2}\pm\mu_I
\;, \\
\omega_4(p)&=&\sqrt{p^2+M^2}\;,
\label{w4sym}
\eqa
Note that all four modes are massive as the $O(2)$ symmetry has been restored.

\subsection{Numerical results}
We next discuss the determination of the parameters in the 
Lagrangian~(\ref{lag}). In the vacuum, we have $\rho_0=\mu_I=0$
and $f_{\pi}=\phi_0$. At the tree-level, we have 
\bqa
\label{msdet}
m^2_{\sigma}&=&m_1^2\;,\\
\label{mpidet}
m_{\pi}^2&=&m_2^2\;,\\
H&=&f_{\pi}m^2_{\pi}\;.
\label{hdet}
\eqa
The mass parameter $m^2$ and the coupling constant $\lambda$ can then 
be expressed in terms of the observable masses of the sigma and the pion:
\bqa
\label{m111}
m^2&=&-{1\over2}(m_{\sigma}^2-3m_{\pi}^2)\;,\\
\lambda&=&N{(m_{\sigma}^2-m_{\pi}^2)\over f_{\pi}^2}\;.
\label{m222}
\eqa
Eqs.~(\ref{hdet})--(\ref{m222}) can now be used to determine the constants
$H$, $m^2$, and $\lambda$, given the values 
$m_{\sigma}=600$MeV, $m_{\pi}=139$MeV,and $f_{\pi}=93$MeV.
In the chiral limit, we find
$m^2=-(300$MeV)$^2$ and $\lambda=83.2466$. At the physical point, we obtain
$m^2=-(230.5$MeV)$^2$ and $\lambda=78.7788$.
Once we take into account quantum fluctuations, Eqs.~(\ref{msdet})
and~(\ref{mpidet}) are modified. After renormalization, we find 
\bqa
m_{\sigma}^2&=&m_2^2
-{\lambda m^2_{\pi}\over16\pi^2}\ln{\Lambda^2e\over m_{\pi}^2}\;.
\;\\
\label{msdet1}
m_{\pi}^2&=&m^2_1
-{\lambda m^2_{\pi}\over16\pi^2}\ln{\Lambda^2e\over m_{\pi}^2}\;.
\label{mpidet1}
\eqa
Note that the renormalization scale $\Lambda$ is
a free parameter and in the chiral limit, it does not enter our expressions. 
At the physical point we chose $\Lambda^2=m_{\pi}^2/e$, so that
the tree-level relations still hold without having to change the parameters
$m^2$ and $\lambda$.

\begin{figure}[htb]
\includegraphics[width=8.8cm]{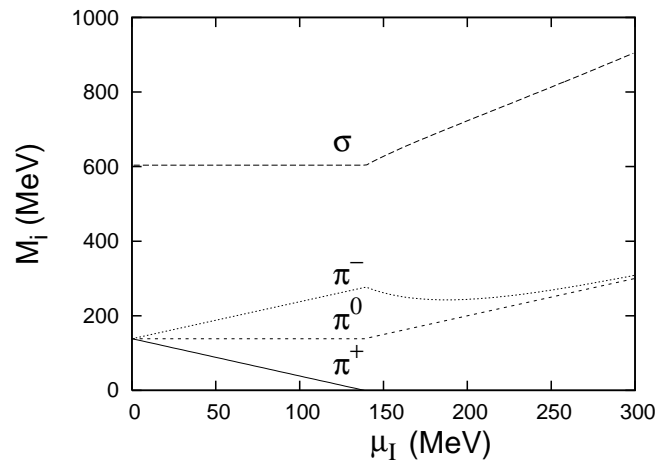}
\caption{
Pion and sigma masses at the physical point
as functions of the isospin chemical potential $\mu_I$ and $T=0$.}
\label{pmass}
\end{figure}

In Fig.~\ref{pmass}, we show the medium dependent
meson masses at the physical point
as functions of the isospin chemical potential at zero temperature. 
For $\mu_I>\mu_{c}=139$ MeV, 
they are given by Eqs.~(\ref{w1br})-(\ref{w4br}), while
for $\mu_I<\mu_{c}$, they are given by Eqs.~(\ref{w1sym})-(\ref{w4sym})
with $p=0$.
For $\mu_I>\mu_{c}$, the mass of $\pi^+$ vanishes.

\begin{figure}[htb]
\includegraphics[width=8.8cm]{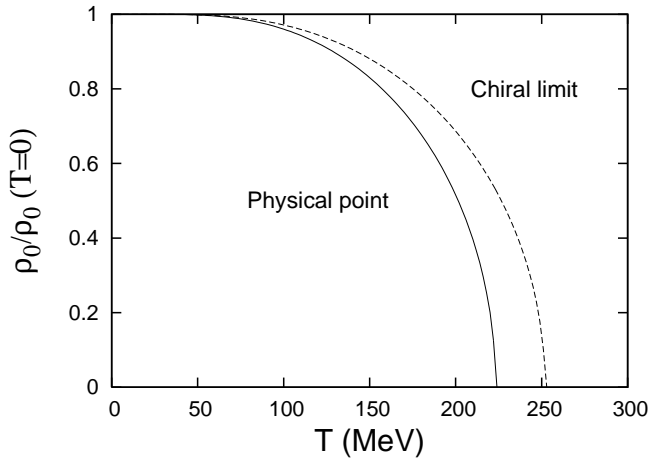}
\caption{
Charged 
pion condensate as a function of temperature normalized to the condensate
at $T=0$. The upper curve is the chiral limit and the lower curve corresponds 
to the physical point.}
\label{fasep0}
\end{figure}

In Fig.~\ref{fasep0}, we show the pion condensate as a function of
temperature normalized to the condensate at $T=0$. In both
the chiral limit and at the physical point, the pion condensate breaks
an $O(2)$ symmetry. From universality arguments, we know this transition
is of second order, which can be seen from Fig.~(\ref{fasep0}).
The critical exponent $\nu={1/2}$, which
is expected from a mean-field calculation. See Ref.~\cite{alford2n} for
a NLO calculation of the critical exponents using the 2PI formalism.

\begin{figure}[htb]
\includegraphics[width=8.8cm]{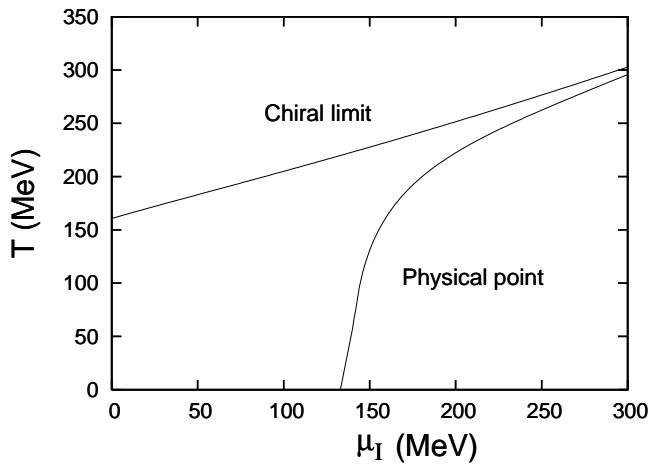}
\caption{
Phase diagram for pion condensation. The upper curve is the chiral limit and 
the lower curve corresponds to the physical point.}
\label{fasep1}
\end{figure}

In Fig.~\ref{fasep1}, we show the phase diagram for pion condensation.
The solid line corresponds to the chiral limit, while the dashed line
corresponds to the physical point. The two curves meet at large $\mu_I$
since the temperature-independent value of $H$ becomes irrelevant.

\begin{figure}[htb]
\includegraphics[width=8.8cm]{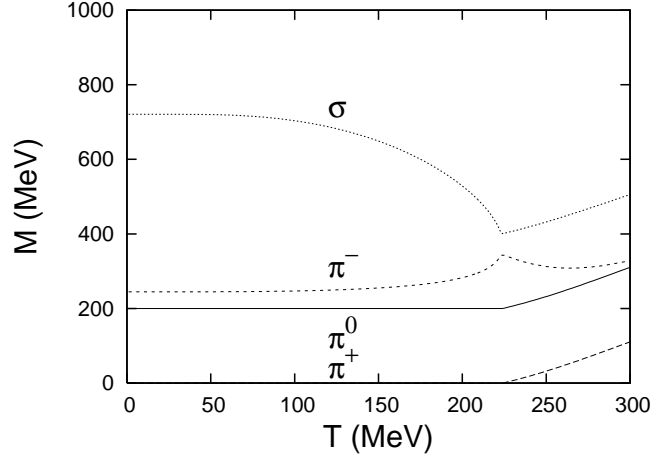}
\caption{Meson masses as functions of temperature at the physical point. 
The isospin chemical potential is
$\mu_I=200$ MeV. $T_c=223$ MeV.}
\label{ftp1}
\end{figure}

In Fig.~\ref{ftp1}, we show the meson masses as functions of temperature
for fixed isospin chemical potential $\mu_I=200$ MeV. 
For $T<T_c$, they are given by Eqs.~(\ref{w1br})-(\ref{w4br}), while
for $T>T_c$, they are given by Eqs.~(\ref{w1sym})-(\ref{w4sym}) with $p=0$.
The critical temperature is $T_c=223$ MeV.
Note that one of the
pions is massless below the critical temperature 

\section{Kaons}
In the color-flavor locked phase of high baryon-density QCD,  
the lightest excitations are the four charged and neutral kaons,
$K^{\pm}$, $K^0$ and $\bar{K}^0$.
The low-energy effective theory is described by the 
Lagrangian~(\ref{lag})
with $H=0$ and $N=2$~\cite{frankie}. 
In this section, we discuss kaon condensation in the presence of
chemical potentials $\mu_Y$ for hypercharge $Y$ and $\mu_I$
for isospin.
The kaons can be written as a complex doublet,
$(\Phi_1,\Phi_2)=(K^0,K^+)$.
We then have the following relation between the various chemical potentials:
\bqa
\mu_I&=&{1\over2}(\mu_2-\mu_1)\;,\\
\mu_Y&=&
\mu_1+\mu_2
\;.
\eqa
We first write the complex fields $\Phi_1$ and $\Phi_2$ in terms of
the charged and neutral kaon condensates $\phi_0$ and $\rho_0$ and quantum 
fluctuating fields $\phi_1-\phi_4$:
\bqa
\Phi_1&=&
{1\over\sqrt{2}}\left(\phi_0+\phi_1+i\phi_2\right)\;,\\
\Phi_2&=&
{1\over\sqrt{2}}\left(\rho_0+\phi_3+i\phi_4\right)
\;.
\eqa
\begin{widetext}
The tree-level propagator is
\bqa
D_0^{-1}(\omega_n,p)=
\left(\begin{array}{cccccc}
\omega_n^2+p^2+m_1^2&-2\mu_1\omega_n&{\lambda\over N}\phi_0\rho_0&0&0&...\vspace{2mm}
\\
2\mu_1\omega_n&\omega_n^2+p^2+m_2^2&0&0&0&...
\\
{\lambda\over N}\phi_0\rho_0&0&\omega_n^2+p^2+m_3^2&-2\mu_2\omega_n&0&...
\\
0&0&2\mu_2\omega_n&\omega_n^2+p^2+m_4^2&0&... \\
...&...&...&...&\omega_n^2+p^2+m^2_5&...
\end{array}\right)\;,
\eqa
\end{widetext}
where the tree-level masses are
\bqa
m_1^2&=&-\mu_1^2+m^2+{3\lambda\over2N}\phi_0^2+{\lambda\over2N}\rho_0^2\;,\\
m_2^2&=&-\mu_1^2+m^2+{\lambda\over2N}\phi_0^2+{\lambda\over2N}\rho_0^2\;, \\
m_3^2&=&-\mu_2^2+m^2+{\lambda\over2N}\phi_0^2+{3\lambda\over2N}\rho_0^2\;,\\
m_4^2&=&-\mu_2^2+m^2+{\lambda\over2N}\phi_0^2+{\lambda\over2N}\rho_0^2\;,\\
m_5^2&=&m^2+{\lambda\over2N}\phi_0^2+{\lambda\over2N}\rho_0^2\;.
\eqa
Note that in the remainder of this section, the mass parameter $m^2$ is 
positive. The 2PI effective potential is given by
\bqa\nonumber
\Gamma[\phi_0,\rho,D]&=&
{1\over2}
\left(m^2-\mu_1^2\right)\phi_0^2
+{1\over2}
\left(m^2-\mu_2^2\right)\rho_0^2
\\&&\nonumber
+{\lambda\over8N}(\phi_0^2+\rho^2)^2 
+{1\over2}{\rm Tr}\ln D^{-1}
\\&&
-{1\over2}{\rm Tr}D_0^{-1}D
+\Phi[D]\;.
\eqa
To leading order in $N$, the contribution to $\Phi[D]$ is again given by the
diagrams in Fig.~\ref{lo}.
The stationarity conditions~(\ref{e1})--(\ref{e2}) 
for the effective potential are 
\bqa\nonumber
0&=&(m^2-\mu_1^2)\phi_0+{\lambda\over2N}\phi_0\left(\phi_0^2+\rho_0^2\right)
\\ &&
+{\lambda}\phi_0\sumint_Q{1\over Q^2+M^2}\;,
\label{min1}
\\ \nonumber
0&=&(m^2-\mu_2^2)\rho_0+{\lambda\over2N}\rho_0\left(\phi_0^2+\rho_0^2\right)
\\ &&
+{\lambda}\rho_0\sumint_Q{1\over Q^2+M^2}\;.
\label{min2}
\eqa
The self-energy to leading order in $1/N$ is again diagonal and 
contributing diagram is shown in Fig.~\ref{self}.
The expression is given by Eq.~(\ref{gappi}).
Using Eq.~(\ref{min1}) for the minimum of the effective potential, we obtain 
\bqa
M^2&=&\mu_1^2\;.
\eqa

We first assume that $\mu_1=\mu_2$. In this case, 
the Lagrangian~(\ref{lag}) has an extended
$SU(2)\times U(1)$ symmetry instead of $O(2)\times O(2)$. 
If $\mu_1^2>m^2$ this 
symmetry is broken down to $U(1)$.
We have complete symmetry between $\phi_0$ and $\rho_0$ and 
it is then possible to rotate e.g. the $\rho_0$ condensate away.
If $\mu_1>\mu_2$ 
(to be specific), $\rho_0=0$ and $\phi_0\neq0$ only if $\mu_1^2>m^2$.
In this case the $O(2)\times O(2)$ 
symmetry is spontaneously broken down to $O(2)$ by the condensate.

In analogy with Eq.~(\ref{g0}), the kaon condensate $\phi_0$ satisfies the
gap equation
\bqa
\phi_0^2&=&{2N\over\lambda}\left(\mu_1^2-m^2\right)
-2\sumint_Q{1\over Q^2+M^2}\;.
\eqa
The phase transition is again of second order. In the weak-coupling limit,
an analytic expression for $T_c$ analogous to Eq.~(\ref{tcw}) is obtained
by the substitution $\mu_I\rightarrow\mu_1$.

In the broken phase, the dispersion relations can be written as
\begin{widetext}
\bqa
\label{w12}
\omega_{1,2}(p)&=&
\sqrt{p^2+2\mu_1^2+{1\over2}(M_1^2+M_2^2)\pm\sqrt{(M_1^2-M_2^2)^2
+8\mu_1^2(2\mu_1^2+M_1^2+M_2^2)+4\mu_1^2p^2}}
\;,\\
\omega_{3,4}(p)&=&
\sqrt{p^2+\mu_1^2}\pm\mu_2\;, \\
\omega_5(p)&=&\sqrt{p^2+\mu_1^2}
\;,
\label{w21}
\eqa
\end{widetext}
where $M^2_i=m^2_i+\Pi_{\rm LO}$.
The neutral kaons $\bar{K^0}$ and $K^0$ are identified with the linear
combinations of the 
fields $\phi_1$ and $\phi_2$, and therefore with
$\omega_1(p)$ and $\omega_2(p)$. Similarly, the charged kaons
$K^-$ and $K^+$ are given by linear combinations of $\phi_3$ and $\phi_4$
and are identified with $\omega_3(p)$ and $\omega_4(p)$.

At the minimum of the effective potential, we have 
$M_1^2=2(\mu_1^2-\tilde{m}^2)$
and $M_2^2=0$, where $\tilde{m}^2=m^2+\Pi_{\rm LO}$.
Eq.~(\ref{w12}) then reduce to
\bqa\nonumber
\omega_{1,2}(p)&=&\sqrt{p^2+3\mu_1^2-\tilde{m}^2\pm
\sqrt{(3\mu_1^2-\tilde{m}^2)^2+4p^2\mu_1^2}}\;.
\\ &&
\eqa
This shows that $\omega_2(p)$ and thus $K^0$
is a massless excitation which is linear for
small $p$ (cf.~(\ref{l1})) and has the quantum numbers of $K^0$. 
In the case where $\mu_1\neq\mu_2$, this is the only Goldstone boson which
corresponds to the broken $O(2)$ symmetry.
In the case where $\mu_1=\mu_2$,
$\omega_4(p)$ is another massless modes,which is quadratic for 
small $p$, which has the quantum numbers of $K^+$.
Thus there are two massless modes
despite the fact that there
are 3 broken 
generators~\cite{shot,igor1,igor2}. The splitting between $K^+$ and 
$K^-$ is due to the fact that the chemical potentials break the 
discrete symmetries $C$, $CP$, and $CPT$~\cite{igor2}.
This result is in agreement with the counting
rules derived by Nielsen and Chadha~\cite{holger}, which state that 
the massless modes with a quadratic dispersion relation must be counted twice.
Finally, we notice that only in the case where $\mu_1\neq\mu_2$, the
condensate implies that the system is superfluid. Whenever the Goldstone
mode with a quadratic dispersion relation is present, the Landau criterion
for superfluidity can never be satisfied~\cite{igor1,igor2}.

We next discuss the system in the normal phase, where $\phi_0=0$.
The dispersion relations reduce to
\bqa
\label{wb11}
\omega_{1,2}(p)&=&\sqrt{p^2+M^2}\pm\mu_1\;,\\
\omega_{3,4}(p)&=&\sqrt{p^2+M^2}\pm\mu_2\;,\\
\omega_{5}(p)&=&\sqrt{p^2+M^2}\;,
\label{wb22}
\eqa
where $M^2$ again satisfies the gap equation~(\ref{selfcon}).
Since the symmetry is restored, all excitations are gapped.

\begin{figure}[htb]
\includegraphics[width=8.8cm]{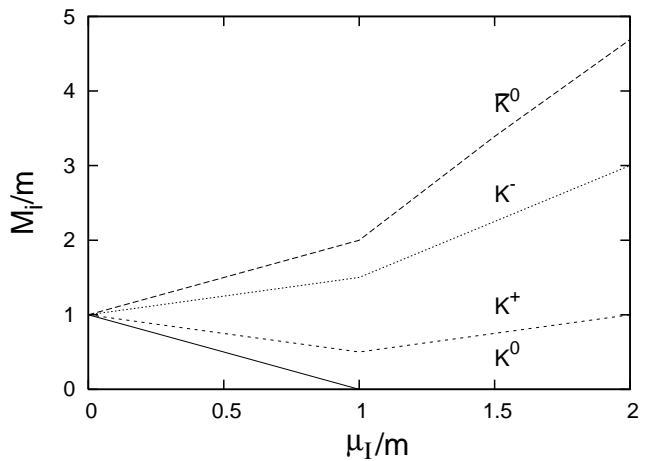}
\caption{
Kaon masses at zero temperature as functions of the chemical potential
$\mu_1$, where $\mu_1>\mu_2$.}
\label{kvasik1}
\end{figure}

In Fig.~\ref{kvasik1}, we show the quasiparticle masses as a function of 
$\mu_1$ at zero temperature. These are given by Eqs.~(\ref{w12})-(\ref{w21})
with $p=0$ in the broken phase and by 
Eqs.~(\ref{wb11})-(\ref{wb22}) with $p=0$ in the symmetric phase.
In the broken phase, the neutral kaon $K^0$ is a  massless Goldstone mode.
We have set $\mu_2=\mu_1/2$ and chosen the renormalization scale
$\Lambda=M$ in the symmetric phase. The phase diagram is the same
as in the pion case with $H\neq0$. 

\section{Summary}
In the present paper, we have calculated to leading order in the 2PI-$1/N$
expansion finite density and finite-temperature effects in
interacting Bose systems with pion and kaon condensation.
In particular, we have obtained the phase diagrams and quasiparticle masses. 
Depending on the magnitude of the chemical potentials, the number and nature
of the Goldstone modes vary. In all cases, our results are in agreement with
general theorems on the occurence of massless 
excitations~\cite{holger,shot,brauner}.
This is in contrast with e.g. the Hartree
approximation, where  the Goldstone theorem is not satisfied.
Furthermore, we have shown that is is possible to renormalize the theory in a 
medium-independent way by eliminating the divergences through a
redefinition of the mass and coupling constant. 


The present work can be extended by calculating the $1/N$ corrections
either in the context of the standard effective action or by using the
2PI effective-action formalism. Using the conventional 1PI effective
action, the calculations would
be similar to those recently carried out in Refs.~\cite{gertie,abw}.
Calculations within the 2PI effective action will be more difficult
as the resulting equations form a 
set of coupled integro-differential equations.
These equations must be solved numerically, which is nontrivial.
However, progress has 
been made recently~\cite{jurgen} 
in massive scalar field theory with vanishing chemical potential.

\section*{Acknowledgment}
The author would like to thank D. Boer and
H. J. Warringa for useful discussion and
suggestions.

\appendix
\renewcommand{\theequation}{\thesection.\arabic{equation}}


\end{document}